\documentclass[aps,prd,twocolumn,groupedaddress,floatfix,showpacs]{revtex4}
\usepackage{mathrsfs}
\usepackage{amsmath}
\usepackage{latexsym}
\usepackage{amsfonts}
\usepackage{graphicx}
\begin{document}
 \title{Constraints on Cosmological Models and Reconstructing the Acceleration History of the Universe
  with Gamma-Ray Burst Distance Indicators}
 \author{Nan~Liang\,}
 \email[\,email address:\ ]{liangn@bnu.edu.cn}
 \affiliation{Department of Astronomy, Beijing Normal University, Beijing 100875, China}
 \author{Puxun~Wu\,}
 \email[\,email address:\ ]{wpx0227@gmail.com}
 \affiliation{Department of Physics and Institute of Physics,\\ Hunan Normal University, Changsha, Hunan 410081, China}
 \author{Shuang~Nan~Zhang}
  \email[\,email address:\ ]{zhangsn@ihep.ac.cn}
 \affiliation{Key Laboratory of Particle Astrophysics, Institute of High Energy Physics,\\Chinese Academy of Sciences, Beijing 100049,
 China \\Physics Department, University of Alabama in Huntsville, Huntsville, Alabama 35899, USA}
\begin{abstract}
Gamma-ray bursts (GRBs) have been regarded as standard candles at
very high redshift for cosmology research. We have proposed a new
method to calibrate GRB distance indicators with Type Ia supernova
(SNe Ia) data in a completely cosmology-independent way to avoid the
circularity problem that had limited the direct use of GRBs to probe
cosmology [N. Liang, W. K. Xiao, Y. Liu, and S. N. Zhang, Astrophys.
J. 685, 354 (2008).]. In this paper, a simple method is provided to
combine GRB data into the joint observational data analysis to
constrain cosmological models; in this method those SNe Ia data
points used for calibrating the GRB data are not used to avoid any
correlation between them. We find that the $\Lambda$CDM model is
consistent with the joint data in the 1-$\sigma$ confidence region,
using the GRB data at high redshift calibrated with the
interpolating method, the Constitution set of SNe Ia, the cosmic
microwave background radiation from Wilkinson Microwave Anisotropy
Probe five year observation, the baryonic acoustic oscillation from
the spectroscopic Sloan Digital Sky Survey Data Release 7 galaxy
sample, the x-ray baryon mass fraction in clusters of galaxies, and
the observational Hubble parameter versus redshift data. Comparing
to the joint constraints with GRBs and without GRBs, we find that
the contribution of GRBs to the joint cosmological constraints is a
slight shift in the confidence regions of cosmological parameters to
better enclose the $\Lambda$CDM model. Finally, we reconstruct the
acceleration history of the Universe up to $z>6$ with the distance
moduli of SNe Ia and GRBs and find some features that deviate from
the $\Lambda$CDM model and seem to favor oscillatory cosmology
models; however further investigations are needed to better
understand the situation.
\end{abstract}
\pacs{98.80.Es,98.80.-k,98.80.Jk} \maketitle

\section{Introduction}
Gamma-ray bursts (GRBs) are the most intense explosions observed so
far and likely to occur in high redshift range. Recently, GRB 090423
has been  observed at a very high redshift, $z>8$ \cite{GRB0904}.
The early universe can be explored by using GRBs at the high
redshift which is hardly achievable by Type Ia supernova (SNe Ia).
In recent years, several GRB luminosity relations between measurable
properties of the prompt gamma-ray emission with the luminosity or
energy have been proposed as distance indicators, many of which are
the two-dimensional (2D) luminosity relations, such as the isotropic
energy ($E_{\rm iso}$) - peak spectral energy ($E_{\rm peak}$)
relation (i.e., the so-called Amati relation) \cite{Amati}, the
luminosity ($L$) - spectral lag ($\tau_{\rm lag}$) relation
\cite{Nor00}, the $L$ - variability ($V$) relation \cite{VL}, the
$L$ - $E_{\rm peak}$ relation \cite{EpL}, the $L$ - minimum rise
time ($\tau_{\rm RT}$) relation \cite{Sch02}, and the
collimation-corrected energy ($E_{\gamma}$) - $E_{\rm peak}$
relation (i.e., the so-called Ghirlanda relation) with small
scattering \cite{GGL04}. In order to reduce the large scattering in
some of these 2D relations, several 3D luminosity relations are also
presented \cite{LZ05,Fir06,Yu09}. Liang and Zhang first proposed the
relation among $E_{\rm iso}$, $E_{\rm peak}$ and $t_{\rm b}$ (i.e.,
the so-called Liang-Zhang relation) \cite{LZ05}, where $t_{\rm b}$
is the break time of the optical afterglow light curves; and Firmani
et al. found the relation among $L$, $E_{\rm peak}$ and $T_{0.45}$
(i.e., the so-called Firmani relation) \cite{Fir06}, where
$T_{0.45}$ is the ``high-signal'' timescale of the prompt emission.
Both of the 3D luminosity relations above could reduce the
scattering in the luminosity or energy \emph{versus} the $E_{\rm
peak}$ relation. More recently, Yu et al. find that, for the 3D
luminosity relations between the luminosity and an energy scale
$E_{\rm peak}$ and a timescale ($\tau_{\rm lag}$ or $\tau_{\rm
RT}$), the intrinsic scattering is considerably smaller than that of
those corresponding 2D luminosity relations \cite{Yu09}. Other GRB
luminosity relations have been proposed in many works
\cite{Hakkila08,Willingale07,Tsutsui09J}. For reviews of GRB
luminosity relations, see e.g. \cite{zhang07,SC07,GGF06,Sch07}.

However, all the luminosity relations of GRBs presented above are
always empirical, but still without solid physical interpretations
\cite{GGF06}. Moreover, the luminosity relations can not be
calibrated with a sufficiently large low-redshift GRB sample.
Therefore previous calibrations have been usually obtained by
assuming a particular cosmological model. Once these GRB luminosity
relations are calibrated, GRB data could be considered as standard
candles at very high redshift for cosmology research
\cite{GGF06,Sch07,Sch03,Bloom03,Dai04,GGLF04,LZ05,Xu05,Fir05,Fir06b,Fir07,Friedman05,Wang06,Cuesta08,BP08,Wri07,Wang07,Gong07,Qi08,Taka03,Berto06}
(see e.g. \cite{GGF06} and \cite{Sch07} for reviews). In 2003,
Schaefer derived the luminosity distances of nine GRBs with known
redshifts by using two GRB luminosity relations ($L$ - $\tau_{\rm
lag}$, $L$ - $V$) to construct the first GRB Hubble diagram
\cite{Sch03}. For 16 GRBs with redshift measurements, Bloom et al.
found a narrow clustering of geometrically-corrected Gamma-ray
energies within the framework of the uniform conical jet model
\cite{Bloom03}. Dai et al.  proposed an approach to consider the
Ghirlanda relation to constrain cosmological parameters and dark
energy \cite{Dai04}. Because of the small scatter, the 3D luminosity
relations have been used for cosmological constraint, including the
Liang-Zhang relation \cite{LZ05}, and the Firmani relation
\cite{Fir06b,Fir07}. In 2007, Schaefer used five 2D relations ($L$ -
$\tau_{\rm lag}$, $L$ - $V$, $L$ - $E_{\rm peak}$, $L$ - $\tau_{\rm
RT}$ and $E_{\gamma}$ - $E_{\rm peak}$) to construct the Hubble
diagram of 69 GRBs \cite{Sch07}. These GRB data have been widely
used to constrain cosmology and dark energy models in many recent
works \cite{Cuesta08,BP08}, and joint constraints by combining these
GRB data with SNe Ia and the other cosmological probes have been
derived in \cite{Wri07,Wang07,Gong07,Qi08}. Instead of a hybrid
sample over the whole redshift range of GRBs, Takahashi et al.
\cite{Taka03} firstly calibrated GRB relations ($L$ - $\tau_{\rm
lag}$, $L$ - $V$) at low redshift where distance-redshift relations
have been already determined from SNe Ia. This method was adopted by
Bertolami and Silva for considering the use of GRBs at $1.5<z<5$
calibrated with the bursts at $z\le1.5$ as distance markers to study
the generalized Chaplygin gas model \cite{Berto06}. All the
calibration methods above carried out have been derived usually from
the $\Lambda$CDM model with particular model parameters according to
the concordance cosmology.

An important point related to the use of GRBs for cosmology is the
dependence on the cosmological model in the calibration of GRB
relations, which had limited the direct use of GRBs for cosmology
research. In order to investigate cosmology, the relations of
standard candles should be calibrated in a cosmology-independent
way; otherwise, the circularity problem cannot be avoided easily
\cite{GGF06}. Recently, the possibility of calibrating GRBs in a
low-dispersion in redshift near a fiducial redshift has been
proposed \cite{Lamb05,Ghir06}, which has been developed further
based on Bayesian theory \cite{LZ06}. However, the GRB sample
available now is far from what is needed to calibrate the relations
in this way \cite{GGF06,LZ06}. Many of the works treat the
circularity problem with statistical approaches
\cite{Sch03,Fir05,Li08,Amati08,WangY08}. A simultaneous fit of the
parameters in the calibration curves and the cosmology is carried
out to find the optimal GRB relation and the optimal cosmological
model in the sense of a minimum scattering in both the luminosity
relations and the Hubble diagram \cite{Sch03}. Firmani et al.
proposed a Bayesian method to get around the circularity problem
\cite{Fir05}. Li et al. presented a global fitting analysis for the
Ghirlanda relation to deal with the problem \cite{Li08}. Amati et
al. used the $E_{\rm iso}$ - $E_{\rm peak}$ relation  to measure the
cosmological parameter by adopting a maximum likelihood approach
\cite{Amati08}. More recently, Wang has shown that the current GRB
data can be summarized by a set of model independent distance
measurements, with negligible loss of information \cite{WangY08},
which is followed by \cite{Qi09,SR09,WangY09}. However, an input
cosmological model is still required in doing the joint fitting, the
circularity problem can not be circumvented completely by means of
these statistical approaches.

In our previous paper \cite{liang08a}, we presented a new method to
calibrate several GRB luminosity relations in a completely
cosmology-independent manner. Our method avoids the circularity
problem more clearly than previous cosmology-dependent calibration
methods. It is obvious that objects at the same redshift should have
the same luminosity distance in any cosmology. Therefore, the
luminosity distance at any redshift in the redshift range of SNe Ia
can be obtained by interpolating (or by other mathematical approach)
directly from the SNe Ia Hubble diagram. Using the interpolation
method, we calibrated seven GRB luminosity relations ($L$ -
$\tau_{\rm lag}$, $L$ - $V$, $L$ - $E_{\rm peak}$, $L$ - $\tau_{\rm
RT}$, $E_{\gamma}$ - $E_{\rm peak}$, $E_{\textrm{iso}}$ - $E_{\rm
peak}$, and $E_{\rm iso}$-$E_{\rm peak}$-$t_{\rm b}$) with the data
compiled by Schaefer \cite{Sch07}. Then if further assuming these
calibrated GRB relations valid for all long GRB data, we can use the
standard Hubble diagram method to constrain the cosmological
parameters from the GRB data at high redshift obtained by utilizing
the relations \cite{liang08a}. These distance data of GRBs are so
far the most cosmology-independent GRB distance indicators.
Following this cosmology-independent GRB calibration method from SN
Ia \cite{liang08a}, the derived GRB data at high redshift range can
be used to constrain cosmological models without circularity problem
\cite{liang08a,Capozziello08,Izzo09,wei09a,wei09b,WLiang09}.
Capozziello and Izzo first used the $E_{\rm iso}$-$E_{\rm
peak}$-$t_{\rm b}$ relation and the $E_{\gamma}$ - $E_{\rm peak}$
relation calibrated with the so-called Liang method to derive the
related cosmography parameters [deceleration parameters ($q$), jerk
parameters ($j$), and snap parameters ($s$) \cite{Visser04}], which
are only related to the derivatives of the scale factor  without any
a priori assumption \cite{Capozziello08}. Wei and Zhang used the
Amati relation calibrated with the interpolation method to
reconstruct the acceleration history of the Universe and constrain
cosmological models \cite{wei09a,wei09b}.

Furthermore, we proposed another approach to calibrate GRB relations
by using an iterative procedure \cite{liang08b}, which is a
nonparametric method in a model independent manner to reconstruct
the luminosity distance at any redshift in the redshift range of SNe
Ia \cite{Shafieloo06,Shafieloo07,Wu08}. Similar to the interpolation
method, Cardone et al. constructed an updated GRBs Hubble diagram on
six 2D relations calibrated by local regression from SNe Ia
\cite{CCD09}. This method and GRB data have been used to
cosmological constraints with SNe Ia in some following works
\cite{Qi09,CDC09}. Kodama et al. presented that the $L$ - $E_{\rm
peak}$ relation can be calibrated with the empirical formula fitted
from the luminosity distance of SNe Ia \cite{kod08}. This method has
been used to constrain cosmological parameters by combining these
GRB data with SNe Ia in a following work \cite{Tsu09a}. However, it
is noted that this calibration procedure depends seriously on the
choice of the formula and various possible formulas can be fitted
from the SNe Ia data that could give different calibration results
of GRBs. As the cosmological constraints from GRBs are sensitive to
GRBs calibration results \cite{WangY08}, the reliability of this
method should be tested carefully. Moreover, as pointed out in
\cite{WangY08}, the GRB luminosity relations which are calibrated by
this way are no longer completely independent of all the SNe Ia data
points; therefore these GRB data can not be used to directly combine
with the whole SNe Ia dataset to constrain cosmological parameters
and dark energy.

It is easy to find that the number of SNe Ia data points that have
been used in the linear interpolating procedure to obtain the GRB
data is relatively small compared to the whole SNe Ia sample. In
order to combine GRB data obtained by our interpolation method into
the joint observational data analysis to constrain cosmological
models, we can exclude the SNe points that have been used in the
interpolating procedure from the SNe Ia sample used to the joint
constraints. In this work, with the updated distance moduli of the
42 GRBs at $z>1.4$ obtained by the interpolating method
\cite{liang08a} from the Union set of 307 SNe Ia \cite{sn307}, we
provide a simple method to ensure that the calibrated GRB data are
independent of the SNe Ia data used in the joint data fitting to
constrain cosmological models, by using the Constitution set of 397
SNe Ia  \cite{sn397}, along with the cosmic microwave background
radiation (CMB) observation  from the five-year Wilkinson Microwave
Anisotropy Probe (WMAP5) result \cite{wmap5}, ($R$, $l_a$,
$\Omega_bh^2$); the baryonic acoustic oscillation (BAO) \cite{bao}
observation from the spectroscopic Sloan Digital Sky Survey (SDSS)
Data Release 7 (DR7) galaxy sample \cite{SDSS7} ($d_{0.2}$,
$d_{0.35}$); the 26 baryon mass fraction in clusters of galaxies
(CBF) from the x-ray gas observation ($f_{\textrm{gas}}$)
\cite{Allen04}; and the 11 Hubble parameter versus redshift data,
$H(z)$ \cite{Simon05,Gazta08}. Our goal is to determine the
contribution of GRBs to the joint cosmological constraints in the
confidence regions of cosmological parameters by comparing to the
joint constraints with GRBs and without GRBs. Finally we also
reconstruct the acceleration history of the Universe up to $z>6$
with the distance moduli of SNe Ia and GRBs.

\section{Observational Data Analysis}

The updated distance moduli of the 42 GRBs at $z>1.4$ are obtained
by the five GRB luminosity relations that are calibrated from the 27
GRBs at $z<1.4$ with the interpolating method \cite{liang08a} using
the Union set of 307 SNe Ia \cite{sn307}. The Union set includes the
Supernova Legacy Survey (SNLS) \cite{Astier05} and the ESSENCE
Survey \cite{WV07}, the extended dataset of distant SNe Ia observed
with the Hubble Space Telescope \cite{Riess07}, and the formerly
observed SNe Ia data. For the SNe data in joint analysis, we use the
new Constitution set of 397 SNe Ia \cite{sn397}, which combine 90
CfA3 SNe Ia sample with the Union set \cite{sn307}. We also consider
the shift parameters set ($R$, $l_a$, $\Omega_bh^2$) \cite{wmap5},
the BAO \cite{bao} parameters set ($d_{0.2}$, $d_{0.35}$)  from SDSS
DR7 \cite{SDSS7}, the 26 $f_{\textrm{gas}}$ data from the x-ray gas
observation in clusters of galaxies \cite{Allen04}, and the 11
$H(z)$ data from the differential ages of passively evolving
galaxies \cite{Simon05} and from the BAO peak position
\cite{Gazta08}.

In our previous paper \cite{liang08a}, we used the data for 69 GRBs
compiled by Schaefer \cite{Sch07} and adopted the data for 192 SNe
Ia compiled by Davis et al.  \cite{Davis07} to calibrate seven GRB
luminosity relations ($L$ - $\tau_{\rm lag}$, $L$ - $V$, $L$ -
$E_{\rm peak}$, $L$ - $\tau_{\rm RT}$,  $E_{\gamma}$ - $E_{\rm
peak}$,  $E_{\textrm{iso}}$ - $E_{\rm peak}$, and $E_{\rm
iso}$-$E_{\rm peak}$-$t_{\rm b}$) at $z\le1.4$ by using the
interpolation method. In this work, we use the Union set of 307 SNe
Ia \cite{sn307} to determine the values of the intercepts and the
slopes of GRBs luminosity relations calibrated with the GRB sample
at $z\le1.4$ by using the linear interpolation method. We do not
include the 90 SNe Ia data from CfA3  \cite{sn397} due to their
extremely low redshift ($z < 0.1$), which would not affect the
calibrated results for GRB luminosity relations at $0.17\le
z\le1.4$. The 2D luminosity relation of GRBs can be generally
written in the form
\begin{eqnarray}
\log y=a+b\log x,
\end{eqnarray}
where $y$ is the luminosity ($L$/$\rm erg\ s^{-1}$) or energy
($E_{\gamma}$/ $\rm erg$); $x$ is the GRB parameters measured in the
rest frame, e.g., $\tau_{\rm lag}(1+z)^{-1}/(0.1\rm \ s)$,
$V(1+z)/0.02$, $E_{\rm peak}(1+z)/(300\ \rm keV)$, $\tau_{\rm
RT}(1+z)^{-1}/\rm (0.1\ s)$, for the corresponding  2D relations. We
adopt the data for these quantities from Ref.  \cite{Sch07}
including five 2D GRB luminosity relations ($L$ - $\tau_{\rm lag}$,
$L$ - $V$, $L$ - $E_{\rm peak}$, $L$ - $\tau_{\rm RT}$ and
$E_{\gamma}$ - $E_{\rm peak}$). For the $E_{\gamma}$ - $E_{\rm
peak}$ relation, in order to calculate the total
collimation-corrected energy $E_{\gamma}$, one needs to know the
beaming factor, $F_{\rm beam}=(1-\cos\theta _{\rm jet})$, where the
value of the jet opening angle $\theta _{\textrm{jet}}$ is related
to the jet break time ($t_{\rm b}$) and the isotropic energy for an
Earth-facing jet, $E_{\gamma ,\rm iso,52}=E_{\gamma ,\rm
iso}/10^{52} \rm erg$ \cite{Sari99}.  When calculating $E_{\gamma
,\rm iso}$, we also use the interpolation method from SNe Ia to
avoid the circularity problem. We determine the values of the
intercept ($a$) and the slope ($b$) with their 1-$\sigma$
uncertainties calibrated with the GRB sample at $z\le1.4$ by using
the linear interpolation methods from the Union set of 307 SNe Ia.

Further assuming that these GRB luminosity relations do not evolve with redshift, we are able to obtain the
luminosity ($L$) or energy ($E_\gamma$) of each burst at high redshift ($z>1.4$). Therefore, the luminosity
distance ($d_L$) can be derived. A distance modulus can be calculated as $\mu=5\log (d_L/\textrm{Mpc}) + 25$.
The weighted average distance modulus from the five relations for each GRB is $ \mu = (\sum_i \mu_{\rm i} /
\sigma_{\mu_{\rm i}}^2)/(\sum_i \sigma_{\mu_{\rm i}}^{-2})$, with its uncertainty $ \sigma_{\mu} = (\sum_i
\sigma_{\mu_{\rm i}}^{-2})^{-1/2}$, where the summations run from 1 to 5 over the five relations with available
data. For more details of the calculation, see \cite{Sch07} and \cite{liang08a}. We have plotted the Hubble
diagram of 397 SNe Ia \cite{sn397} which combine 90 CfA3 SNe Ia sample with the Union set \cite{sn307} and the
69 GRBs obtained  using the interpolation methods in Fig 1. The distance moduli of the 27 GRBs at $z\le1.4$ are
obtained by using the linear interpolation method directly from the Union SNe data. The 42 GRB data at $z>1.4$
are obtained by utilizing the five relations calibrated with the sample at $z\le1.4$ using the interpolation
method.

 \begin{figure}
\includegraphics[angle=0,width=0.45\textwidth]{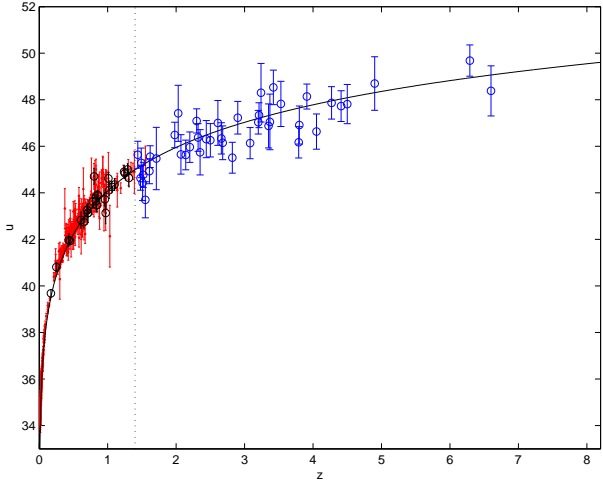}
\caption{Hubble Diagram of 397 SNe Ia (\emph{red dots}) and the 69
GRBs (\emph{circles}) obtained using the interpolation methods. The
27 GRBs at $z\le1.4$ are obtained by linear interpolating from SNe
Ia data (\emph{black circles}), and the 42 GRBs at $z>1.4$
(\emph{blue circles}) are obtained with the five relations
calibrated with the sample at $z\le1.4$ using the interpolation
method. The curve is the theoretical distance modulus in the
concordance model ($w= -1$, $\Omega_{\rm M0}= 0.27$,
$\Omega_{\Lambda}= 0.73$), and the vertical dotted line represents
$z=1.4$. } \label{fHD}
\end{figure}

Constraints from the SNe Ia data and the GRB data can be obtained by fitting the distance moduli $\mu(z)$. A
distance modulus can be calculated as
\begin{eqnarray}\label{mu}
\mu=5\log \frac{d_L}{\textrm{Mpc}} + 25=5\log_{10}D_L-\mu_0,
\end{eqnarray}
where $\mu_0=5\log_{10}h+42.38$, $h=H_0/(100{\rm km/s/Mpc})$, $H_0$ is the Hubble constant, and the luminosity
distance $D_L$ can be calculated by
\begin{eqnarray}\label{dLH}
D_L\equiv
H_0d_L=(1+z)\Omega_\textrm{k}^{-1/2}\textrm{sinn}\bigg[\Omega_\textrm{k}^{1/2}\int_0^z\frac{dz'}{E(z')}\bigg],
\end{eqnarray}
where $\Omega _{\rm k}=1-\Omega _{\rm M0}-\Omega _{\textrm{DE}}$,
$\rm{ sinn}$$(x)$ is $\rm sinh$ for $\Omega _{\rm k}>0$, $\rm sin$
for $\Omega _{\rm k}<0$, and $x$ for $\Omega _{\rm k}=0$; and $E(z)$
can be given by
\begin{eqnarray}\label{Ez}
E(z)&\equiv&\frac{H}{H_0}\\
\nonumber &=&[\Omega_{\textrm{M0}}(1+z)^3+\Omega_{\textrm{DE}}\exp\bigg[3\int_0^z\frac{1+w(z')}{1+z'}dz'\bigg]\\
\nonumber & &+\Omega_\textrm{k}(1+z)^2]^{1/2},
\end{eqnarray}
where the equation of state (\emph{EoS}) of dark energy $w(z)$ is
determined by the choice of the specific dark energy model. The
$\chi^2$ value of the observed distance moduli can be calculated by
\begin{eqnarray}\label{chi2}
\chi^2_{\mu}=\sum_{i=1}^{N}\frac{[\mu_{\textrm{obs}}(z_i)-\mu(z_i)]^2}
{\sigma_{\mu,i}^2},
\end{eqnarray}
where $\mu _{\rm obs}(z_i)$ is the observed distance modulus for the SNe Ia and/or GRBs at redshift ~$z_i$~ with
its error $\sigma_{\mu_{\rm i}}$; $\mu(z_i)$ is the theoretical value of
distance modulus from a dark energy model which can be calculated 
from  Eqs. -(4). The nuisance parameter $\mu_0$ (or $h$) can be
obtained by marginalizing the likelihood functions  of over $\mu_0$
(or $h$) for all values \cite{Dai04} or treated by following an
effective approach \cite{NP05} to expand the $\chi^2_{\mu}$ to
$\chi^2_{\mu}(\mu_0) = A\mu_0^2  - 2B\mu_0 + C $, with
$A=\sum{1}/{\sigma_{\mu_{i}}^2}$,
$B=\sum{[\mu_{\textrm{obs}}(z_i)-5\log_{10}D_L]}/{\sigma_{\mu_{i}}^2}$,
and
$C=\sum{[\mu_{\textrm{obs}}(z_i)-5\log_{10}D_L]^2}/{\sigma_{\mu_{i}}^2}$.
Thus the $\chi^2_{\mu}$ has a minimum value for $\mu_0={B}/{A}$ at
${\hat\chi}^2_{\mu}$ which given by
\begin{eqnarray}
{\hat\chi}^2_{\mu}=C- \frac{B^2}{A}.
\end{eqnarray}
Therefore we can  minimize ${\hat\chi}^2_{\mu}$ instead of
minimizing $\chi^2_{\mu}$.

For the CMB observation from the WMAP5 results \cite{wmap5}, the two
shift parameters $R$ and $l_a$, together with the  baryon density
fraction of the Universe ($\omega_b=\Omega_bh^2$) can provide an
efficient summary of CMB data to constrain cosmological models. The
shift parameter $R$ can be expressed as
\begin{equation}
R=\Omega_{\textrm{M0}}^{1/2}\Omega_\textrm{k}^{-1/2}\textrm{sinn}\bigg[\Omega_\textrm{k}^{1/2}\int_0^{z_{\textrm{rec}}}\frac{dz}{E(z)}\bigg],
\end{equation}
where $z_{rec}$ is the redshift of recombination which is given by
\cite{Hu1996},
$z_{\textrm{rec}}=1048[1+0.00124\omega_b^{-0.738}(1+g_{1}(\Omega_{\textrm{M0}}h^2)^{g_2})]$,
$g_1=0.0783\omega_b^{-0.238}(1+39.5\omega_b^{-0.763})^{-1}$ and
$g_2=0.560(1+21.1\omega_b^{1.81})^{-1}$. The shift parameter $l_a$
can be expressed as
\begin{equation}
l_a=\pi\frac{\Omega_\textrm{k}^{-1/2}\textrm{sinn}[\Omega_\textrm{k}^{1/2}\int_0^{z_{\textrm{rec}}}\frac{dz}{E(z)}]/H_0}{r_s(z_{\textrm{rec}})},
\end{equation}
where $r_s(z_{\textrm{rec}})$ is the comoving sound horizon at
photo-decoupling epoch,
\begin{eqnarray}\label{cal_rs}
r_s(z_{\textrm{rec}})&=&\frac{1}{H_0}\int_{z_{\textrm{rec}}}^{\infty}\frac{c_s(z)}{E(z)}dz\\
\nonumber&=&a_{\textrm{rec}}\int_0^{a_{\textrm{rec}}}\frac{c_s}{\Omega_{\textrm{M0}}^{1/2}}
\bigg[\frac{\Omega_rh^2}{\Omega_{\textrm{M0}}h^2}+a\bigg]^{-1/2}da~,
\end{eqnarray}
with the sound speed $c_s = (1/\sqrt3)[1 +
a(31500\omega_b(T_{\textrm{CMB}}/2.7 \textrm{ K})^{-4})]^{-1/2}$
\cite{WangY09}, and $T_{\textrm{CMB}} = 2.725 \textrm{ K}$
\cite{wmap5}. From the WMAP5 measurement, the best-fit values of
($R$, $l_a$, $100\omega_b$) for a flat prior are \cite{wmap5}
\begin{eqnarray}
\hspace{-.5cm}\bar{\textbf{P}}_{\textrm{CMB}} &=&
\left(\begin{array}{c}
{\bar R} \\
{\bar l_a}\\
{100\omega_b}\end{array}
  \right)=
  \left(\begin{array}{c}
1.710\pm 0.019 \\
302.10 \pm 0.86\\
2.2765 \pm 0.0596 \end{array}
  \right).
 \end{eqnarray}
The $\chi^2$ value of the CMB observation from WMAP5 can be
expressed as
\begin{eqnarray}
\chi^2_{\textrm{CMB}}=\Delta \textbf{P}_{\textrm{CMB}}^T{\bf
C_{\textrm{CMB}}}^{-1}\Delta\textbf{P}_{\textrm{CMB}},
\end{eqnarray}
where
\begin{eqnarray}
\Delta\bf{P_{\textrm{CMB}}} &=& \left(\begin{array}{c}
R - 1.710 \\
l_a-302.10\\
100\omega_b-2.2765\end{array}
  \right),
\end{eqnarray}
and the corresponding inverse  covariance matrix is \cite{wmap5}
\begin{eqnarray}
\hspace{-.5cm} {\bf C_{\textrm{CMB}}}^{-1}=\left(
\begin{array}{ccc}
2809.73& -0.133381& 158.356\\
-0.133381& 2.21908& 19.7195\\
158.356& 19.7195& 465.728
\end{array}
\right).
\end{eqnarray}
It is noted that we use the priors following \cite{Sanchez09},
$\Omega_b = 0.022765/h^2$ and $h = 0.705$ \cite{wmap5}, when
calculating the value of $\chi^2_{\textrm{CMB}}$.

For the BAO observation \cite{bao}, from the SDSS DR7 galaxy sample
\cite{SDSS7}, we use the measurement of $d_z$ at $z=0.2$ and
$z=0.35$, where $d_z$ can be expressed as \cite{SDSS7}
\begin{equation}
d_z=\frac{r_s(z_d)}{D_V(z_{\textrm{BAO}})}~,
\end{equation}
where $z_d$ is the drag epoch at which baryons were released from
photons which is given by \cite{bao1998},
$z_{d}=1291(\Omega_{\textrm{M0}}h^2)^{0.251}[(1+b_{1}\omega_{b}^{b_2})]/[1+0.659(\Omega_\textrm{M0}h^2)^{0.828}]$,
$b_1=0.313(\Omega_{\textrm{M0}}h^2)^{-0.419}[1+0.607(\Omega_{\textrm{M0}}h^2)^{0.674}]^{-1}$,
and $b_2=0.238(\Omega_{\textrm{M0}}h^2)^{0.223}$, and $D_V$ can be
given by \cite{bao}
\begin{equation} D_V(z_{\textrm{BAO}})=\frac{1}{H_0}\big
[\frac{z_{\textrm{BAO}}}{E(z_{\textrm{BAO}})}\big(\int_0^{z_{\textrm{BAO}}}\frac{dz}{E(z)}\big
)^2\big]^{1/3}~.
\end{equation}
From the SDSS DR7 measurement, the best-fit values are \cite{SDSS7}
\begin{eqnarray}
\hspace{-.5cm}\bar{\textbf{P}}_{\textrm{BAO}} &=&
\left(\begin{array}{c}
{\bar d_{0.2}} \\
{\bar d_{0.35}}\\
\end{array}
  \right)=
  \left(\begin{array}{c}
0.1905\pm0.0061\\
0.1097\pm0.0036\\
\end{array}
  \right).
 \end{eqnarray}
The $\chi^2$ value of the BAO observation from SDSS DR7 can be
expressed as \cite{SDSS7}
\begin{eqnarray}
\chi^2_{\textrm{BAO}}=\Delta \textbf{P}_{\textrm{BAO}}^T{\bf
C_{\textrm{BAO}}}^{-1}\Delta\textbf{P}_{\textrm{BAO}},
\end{eqnarray}
where
\begin{eqnarray}
\Delta\bf{P_{\textrm{BAO}}} &=& \left(\begin{array}{c}
d_{0.2} - 0.1905 \\
d_{0.35} - 0.1097 \\
\end{array}
  \right),
\end{eqnarray}
and the corresponding inverse  covariance matrix is \cite{SDSS7}
\begin{eqnarray}
\hspace{-.5cm} {\bf C_{\textrm{BAO}}}^{-1}=\left(
\begin{array}{ccc}
30124& -17227\\
-17227& 86977\\
\end{array}
\right).
\end{eqnarray}

The  baryon  mass fraction  in clusters of galaxies from the x-ray
gas observation can be used to constrain cosmological parameters on
the assumption that the gas mass fraction in clusters is a constant
and thus independent of redshift. The baryon gas mass fraction
$f_{\textrm{gas}}$ can be presented as \cite{Allen04}
\begin{equation}
f_{\textrm{gas}}(z)=\lambda[\frac{d_A^{\textrm{SCDM}}(z)}{d_A(z)}]^{2/3},
\end{equation}
where $\lambda=[b\Omega_\textrm{b}(2h)^{3/2}]/[(1+a)\Omega_\textrm{\textrm{M0}}]$, $a=0.19\sqrt{h}$, $b$ is a
bias factor motivated by gas dynamical simulations, and $d_A\equiv d_L/(1+z)^2$ is the theoretical value of the
angular diameter distance from cosmological models, $d_A^{\textrm{SCDM}}$ is the angular diameter distance
corresponding to the standard cold dark matter (SCDM) universe ($\Omega_{\textrm{M0}} = 1$ for a flat universe).
Here we adopt the usually used 26 observational $f_{\textrm{gas}}$ data \cite{Allen04} to constrain cosmological
models. The $\chi^2$ value of cluster's baryon gas mass fraction  is
\begin{eqnarray}
\chi^2_{\textrm{CBF}}=\sum_{i=1}^{N=26}\frac{[f_{\textrm{gas}}^{\textrm{obs}}(z_i)-f_{\textrm{gas}}(z_i)]^2}
{\sigma_{f_{\textrm{gas}},i}^2}.
\end{eqnarray}
Following \cite{Nesseris07,Wu07}, we treat $\lambda$ as a nuisance
parameter to expand the $\chi^2_{\textrm{CBF}}$   to
$\chi^2_{\textrm{CBF}}(\lambda) = A\lambda^2 - 2B\lambda + C$, with
$A=\sum[{\tilde{f}_{\textrm{gas},i}}/\sigma_{f_{\textrm{gas}},i}]^2$,
$B=\sum{[\tilde{f}_{\textrm{gas},i}f_{\textrm{gas},i}]}/{\sigma_{f_{\textrm{gas}},i}^2}$,
and $C=\sum{[f_{\textrm{gas},i}}/\sigma_{f_{\textrm{gas}},i}]^2$,
where
${\tilde{f}_{\textrm{gas},i}}=[{d_A^{\textrm{SCDM}}(z)}/{d_A(z)}]^{2/3}$.
Thus the $\chi^2_{\textrm{CBF}}$ has a minimum value  at
${\hat\chi}^2_{\textrm{CBF}}$ which is given by
\begin{eqnarray}
{\hat\chi}^2_{\textrm{CBF}}=C- \frac{B^2}{A}.
\end{eqnarray}

The Hubble parameter $H(z)$ can be derived from the derivative of
redshift with respect to the cosmic time,
\begin{equation}
H(z)=-\frac{1}{1+z}\frac{dz}{dt}.
\end{equation}
From the Gemini Deep Deep Survey (GDDS) \cite{Abraham04}
observations of differential ages of passively evolving galaxies and
other archival data  \cite{Nolan03}, the $H(z)$ data at nine
different redshifts ($0.09\leq z\leq1.75$) have been obtained
\cite{Simon05}. Recently, using the BAO peak position as a standard
ruler in the radial direction, $H=83.2\pm 2.1 {\rm km/s/Mpc}$ at
$z=0.24$, and $H=90.3\pm 2.5 {\rm km/s/Mpc}$ at $z=0.43$ have been
obtained \cite{Gazta08}. To constrain cosmological models, the
$\chi^2$ value of the 11 $H(z)$ data is
\begin{eqnarray}
\chi^2_{H}=\sum_{i=1}^{N=11}\frac{[H_{\textrm{obs}}(z_i)-H(z_i)]^2}
{\sigma_{H,i}^2}.
\end{eqnarray}
The nuisance parameter $H_0$ is also marginalized following the
procedure used in calculating $\hat\chi^2_{\mu}$ and
$\hat\chi^2_{\textrm{CBF}}$ \cite{NP05,Nesseris07,Wu07}.

\section{Constraints on cosmological Models from the joint data with GRBs}
The distance moduli of the 42 GRBs at $z>1.4$ are obtained by the
five GRB luminosity relations that are calibrated from the 27 GRBs
at $z<1.4$ with the interpolating method using directly the distance
moduli of adjacent SNe Ia. In the interpolating procedure to obtain
the distance moduli of the 27 GRBs at $z<1.4$, we have used only 40
SNe Ia data points from the Union set of 307 SNe Ia. In order to
combine GRB data into the joint observational data analysis to
constrain the dark energy models, we exclude the 40 SNe points from
the SNe Ia sample used to the joint constrains. Therefore the
remaining SNe Ia data points are completely independent of the
distance moduli of the 42 GRBs at $z>1.4$. Those excluded 40 SNe are
listed in the Appendix. Since the reduced 357 SNe Ia, 42 GRBs, CMB,
BAO, as well as CBF and $H(z)$ are all effectively independent, we
can combine the results by simply multiplying the likelihood
functions. Thus the cosmological parameters can be fitted with the
combined observable data by the minimum $\chi^2$ method. The total
$\chi^2$ with the SNe + GRBs + CMB + BAO + CBF + $H(z)$ dataset is
\begin{equation}
\chi^2=\hat\chi^2_{\mu,\{\rm{SN+GRB}\}}+\chi^2_{\textrm{CMB}}+\chi^2_{\textrm{BAO}}+{\hat\chi}^2_{\textrm{CBF}}+\hat{\chi}^2_{H}.
\end{equation}
The best-fit values for these parameters can be determined by
minimizing the total $\chi^2$. For comparison, SNe + CMB + BAO + CBF
+ $H(z)$ without GRBs have been used to show the contribution of
GRBs to the joint cosmological constraints, and we also consider the
joint constraints with 397 SNe + CMB + BAO + CBF + $H(z)$. In
addition, some different data set such as SNe + GRBs + CMB + BAO,
SNe + GRBs + CMB, and SNe + GRBs + BAO  have also been used in the
cosmological constraints.

We use  the Akaike information criterion (AIC) \cite{Akaike1974} and the Bayesian information criterion (BIC,
the so-called Schwarz information criterion) \cite{Schwarz1978} to select the best-fit models. Liddle examined
the use of information criteria in the context of cosmological observations \cite{Liddle2004}.  The AIC is
defined as \cite{Akaike1974} $\textrm{AIC}=-2\ln\mathscr{L}_{\textrm{max}}+2k$, where
$\mathscr{L}_{\textrm{max}}$ is the maximum likelihood, and $k$ the number of parameters of the model. Models
with too few parameters give a poor fit to the data and hence have a low log-likelihood, while those with too
many are penalized by the second term. The best model is the model which minimizes the AIC \cite{Liddle2004}.
The BIC can be defined as \cite{Schwarz1978} $\textrm{BIC}=-2\ln\mathscr{L}_{\textrm{max}}+k\ln N$, where $N$ is
the number of data points used in the fit.  Note that for Gaussian errors, $\chi^2_{\textrm{min}} =
-2\ln\mathscr{L}_{\textrm{max}}$, and the difference in BIC can be simplified to
$\textrm{BIC}=\Delta\chi^2_{\textrm{min}}+\Delta k\ln N$. The AIC gives results similar to the BIC approach,
although the AIC is not strict enough on models with extra parameters for any reasonably sized data set ($\ln N
> 2$). Therefore, for comparing cosmological models from the joint data set, we only compare $\Delta$BIC
measured with respect to the best model.  A difference in BIC  of 2 is considered positive evidence against the
model with the higher BIC, while a $\Delta$BIC of 6 is considered strong evidence \cite{Liddle2004}.

The combined data sets  are used  to constrain cosmological
parameters and dark energy. Here  we consider three cosmological
models, the $\Lambda$CDM model with dark energy \textsl{EoS}
$w\equiv-1$, the $w$CDM model with a constant \emph{EoS}, and the
Chevallier-Polarski-Linder (CPL) model in which dark energy with a
parameterization \textsl{EoS}
as \cite{CPL} 
\begin{equation}w(z)=w_0+w_a(1-a)=w_0+w_a\frac{z}{1+z}.\end{equation}
For the $\Lambda$CDM model, Eq.~(\ref{Ez}) becomes
\begin{equation}
E(z) = \Big[\Omega_{\textrm{M0}} (1+z)^3 + \Omega_{\Lambda} -
\Omega_\textrm{k}(1+z)^2\Big]^{1/2},
\end{equation}
where $\Omega_{\textrm{M0}}+\Omega_{\Lambda}+\Omega_\textrm{k}=1$.
For the $w$CDM model with a constant \emph{EoS} for a flat universe
prior,
\begin{eqnarray}
E(z) = \Big[\Omega_{\textrm{M0}} (1+z)^3 +(1-\Omega_{\textrm{M0}})
(1+z)^{3(1+w_0)}\Big]^{1/2},
\end{eqnarray}
and considering $w(z)$ as CPL parameterization model for a flat
universe prior, Eq.~(\ref{Ez}) becomes
\begin{eqnarray}
E(z) &=& \Big[\Omega_{\textrm{M0}}(1+z)^3+(1 -
\Omega_{\textrm{M0}})\\\nonumber & & (1+z)^{3 (1 + w_0 + w_a)} e^{-
3 w_a \frac{z}{1+z}}    \Big]^{1/2}.
\end{eqnarray}

 \begin{table*}[t!]
 \begin{center}{\scriptsize
 \begin{tabular}{c|c|c|c|c} \hline\hline
                         \ \ & $\Lambda$CDM model (with GRBs)  &$\Lambda$CDM model (without GRBs)     &$w$CDM model (with GRBs)          &$w$CDM model (without GRBs) \\ \hline
  $\Omega_{\textrm{M0}}$\ \ & \ \ $0.275^{+0.016}_{-0.015}$\ \ & \ \ $0.270^{+0.016}_{-0.015}$\ \     & \ \ $0.269_{-0.014}^{+0.013}$\ \ & \ \ $0.267_{-0.014}^{+0.013}$\ \ \\
  $\Omega_{\Lambda}    $\ \ & \ \ $0.723^{+0.017}_{-0.016}$\ \ & \ \ $0.730^{+0.017}_{-0.017}$\ \     & \ \ $\Omega_{\Lambda}\equiv1-\Omega_{\textrm{M0}}$ \ \ & \ \ $\Omega_{\Lambda}\equiv1-\Omega_{\textrm{M0}}$\ \ \\
  $w$                   \ \ & \ \ $w\equiv-1$\ \               & \ \ $w\equiv-1$\ \                   & \ \ $-0.99_{-0.07}^{+0.07}$\ \   & \ \ $-0.98_{-0.07}^{+0.07}$\ \ \\
  $\chi_{\rm min}^2$    \ \ & \ \ $494.476$\ \                 & \ \ $449.001$\ \                     & \ \ $494.532$\ \                 & \ \ $448.808$\ \ \\
  $\chi_{\rm min}^2/\textrm{dof}$\ \ & \ \ $  1.129$\ \                 & \ \ $1.134$\ \                       & \ \ $1.129$\ \                   & \ \ $1.133$\ \ \\
  $\textrm{AIC}$     \ \ & \ \ $500.476$\ \                    & \ \ $454.001$\ \                     & \ \ $500.532$\ \                 & \ \ $454.808$\ \ \\
  $\textrm{BIC}$     \ \ & \ \ $512.736$\ \                    & \ \ $466.960$\ \                     & \ \ $512.792$\ \                & \ \ $466.767$\ \ \\
  \hline\hline
 \end{tabular} }
 \end{center}
 \caption{\label{tab1} The best-fit value of $\Omega_{\textrm{M0}}$, $\Omega_{\Lambda}$, and $w$  with
$1\sigma$ uncertainties, and $\chi_{\rm min}^2$, $\chi_{\rm
min}^2/\textrm{dof}$, as well as $\textrm{AIC}$, $\textrm{BIC}$  for
the $\Lambda$CDM model and for the flat $w$CDM model with
SNe+GRBs+CMB+BAO+CBF+$H(z)$ (with GRBs) and SNe+CMB+BAO+CBF+$H(z)$
(without GRBs). }
 \end{table*}

In Fig. 2, we show the joint confidence regions  in the
($\Omega_{\rm M0}, \Omega_{\Lambda}$) plane    for the $\Lambda$CDM
model. With 357SNe + GRBs + CMB + BAO + CBF + $H(z)$, the 1-$\sigma$
confidence region for $(\Omega_{\rm M0},\Omega_{\Lambda})$ of the
$\Lambda$CDM model are $(\Omega_{\rm
M0},\Omega_{\Lambda})=(0.275^{+0.016}_{-0.015},0.723^{+0.017}_{-0.016})$,
with $\chi_{\rm min}^2 =494.476$ for 438 degrees of freedom. For
comparison, fitting results from the joint data  without GRBs are
also given in Fig. 2. With 357SNe + CMB + BAO + CBF + $H(z)$, the
best-fit values are $(\Omega_{\rm
M0},\Omega_{\Lambda})=(0.270^{+0.016}_{-0.015},0.730^{+0.017}_{-0.017})$
with $\chi_{\rm min}^2 =449.001$ for 396 degrees of freedom. We
present the best-fit value of $\Omega_{\textrm{M0}}$,
$\Omega_{\Lambda}$ with $1\sigma$ uncertainties, and $\chi_{\rm
min}^2$, $\chi_{\rm min}^2/\textrm{dof}$, as well as $\textrm{AIC}$,
$\textrm{BIC}$  for the $\Lambda$CDM model in Table 1.

From comparing to the joint constraints with GRBs and without GRBs,
we can see that the contribution of GRBs to the joint cosmological
constraints is a slight shift between the best-fit values near the
line that represents a flat universe, towards a higher matter
density Universe by $\Delta\Omega_{\rm M0}=0.005$, compared to the
joint constraints without GRBs. It is noted that the obtained errors
for these parameters with and without GRBs are essentially
unchanged, because the number of SNe Ia data points dropped is
similar to the number of GRBs when including GRBs in the joint
fitting. This model has the lowest BIC compared to other models
tested from the joint data [357SNe + CMB + BAO + CBF + $H(z)$]  with
GRBs, so $\Delta$BIC are measured with respect to this model.
 \begin{center}
 \begin{figure}[tbhp]
 \centering
 \includegraphics[width=0.45\textwidth]{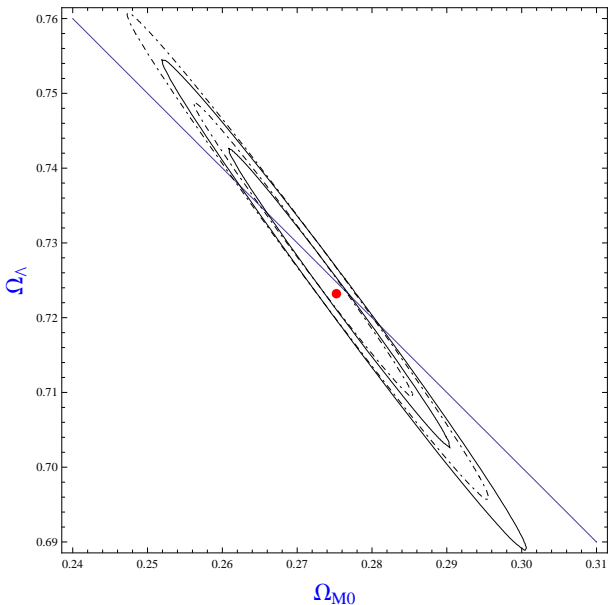}\hfill
 \caption{\label{flcdm}
The joint confidence regions in the  ($\Omega_{\rm M0},
\Omega_{\Lambda}$) plane  for the $\Lambda$CDM model with
SNe+GRBs+CMB+BAO+CBF+$H(z)$. The contours correspond to 1-$\sigma$
and 2-$\sigma$ confidence regions, and the red point is the best-fit
value [SNe+GRBs+CMB+BAO+CBF+$H(z)$]. The results for the cases with
and without GRBs are indicated by the solid lines and the dot-dashed
lines, respectively. The blue line represents a flat universe
($\Omega_{\rm M0}+\Omega_{\Lambda}=1$).}
 \end{figure}
 \end{center}
 \begin{center}
 \begin{figure}[tbhp]
 \centering
 \includegraphics[width=0.45\textwidth]{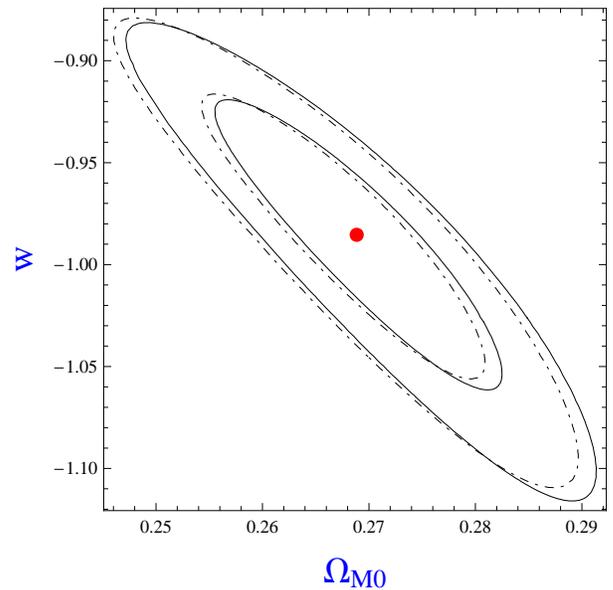}\hfill
 \caption{\label{fwcdm}
The joint confidence regions in the ($\Omega_{\rm M0}, w$) plane for
the dark energy model with a constant $w$ for a flat universe with
 SNe+GRBs+CMB+BAO+CBF+$H(z)$. The contours correspond
to 1-$\sigma$ and  2-$\sigma$ confidence regions, and the red point
is the best-fit value [SNe+GRBs+CMB+BAO+CBF+$H(z)$]. The results for
the cases with and without GRBs are indicated by the solid lines and
the dot-dashed lines, respectively.}
 \end{figure}
 \end{center}

Figure 3 shows the joint confidence regions in the ($\Omega_{\rm
M0}, w$) plane  for the $w$CDM model with a constant \emph{EoS} for
a flat universe.  With 357SNe + GRBs + CMB + BAO + CBF + $H(z)$, the
1-$\sigma$ confidence region for $(\Omega_{\rm M0}, w)$ of the flat
$w$CDM model  are $(\Omega_{\rm
M},w)=(0.269_{-0.014}^{+0.013},-0.99_{-0.07}^{+0.07})$, with
$\chi_{\rm min}^2 = 494.532$ for 438 degrees of freedom. For
comparison, fitting results from the joint data without GRBs are
also given in Fig. 3. With 357SNe + CMB + BAO + CBF + $H(z)$, the
best-fit values are $(\Omega_{\rm
M0},w)=(0.267_{-0.014}^{+0.013},-0.98_{-0.07}^{+0.07})$, with
$\chi_{\rm min}^2 =448.808$ for 396 degrees of freedom. We present
the best-fit value of $\Omega_{\textrm{M0}}$, $\Omega_{\Lambda}$
with $1\sigma$ uncertainties, and $\chi_{\rm min}^2$, $\chi_{\rm
min}^2/\textrm{dof}$, as well as $\textrm{AIC}$, $\textrm{BIC}$  for
the flat $w$CDM model in Table 1.

Comparing to the joint constraints with GRBs and without GRBs, we
can see that the contribution of GRBs to the joint cosmological
constraints is a slight shift that adding the best-fit value of
$\Omega_{\rm M0}$ to $0.002$, and subtracting the best-fit value of
$w$ to $-0.01$ to enclose the $\Lambda$CDM model ($w=-1$). From the
joint data (357SNe + CMB + BAO + CBF + $H(z)$) with GRBs for the
flat $w$CDM model, we obtain $\Delta\textrm{BIC}=0.056$ with respect
to the $\Lambda$CDM model. This has the lowest BIC compared to other
models tested from the joint data [357SNe + CMB + BAO + CBF +
$H(z)$] without GRBs, so $\Delta$BIC are measured with respect to
this model. Therefore, for the $\Lambda$CDM from the joint data
without GRBs , we obtain $\Delta\textrm{BIC}=0.193$  with respect to
the flat $w$CDM model.

 \begin{table*}[t!]
 \begin{center}{\scriptsize
 \begin{tabular}{c|c|c|c|c|c|c} \hline\hline
                       \ \ & 357SNe+GRB+others        &      357SNe+others             & 397SNe+others                  & 357SNe+GRB+C+B           & 357SNe+GRB+C                   & 357SNe+GRB+B\\ \hline
 $w_0$           \ \ & \ \ $-0.98_{-0.15}^{+0.16}$\ \ & \ \ $-0.95_{-0.17}^{+0.16}$\ \ & \ \ $-0.99_{-0.15}^{+0.16}$\ \ & \ \ $-0.96_{-0.16}^{+0.16}$\ \ & \ \ $-0.90_{-0.17}^{+0.17}$\ \ & \ \ $-1.15^{+0.33}_{-0.12}$\ \ \\
 $w_a$           \ \ & \ \ $-0.02^{+0.47}_{-0.60}$\ \ & \ \ $-0.14^{+0.50}_{-0.65}$\ \ & \ \ $0.00^{+0.50}_{-0.62}$\ \  & \ \ $-0.09^{+0.57}_{-0.66}$\ \ & \ \ $-0.21^{+0.59}_{-0.67}$\ \ & \ \  $1.06^{+0.18}_{-2.36}$\ \ \\
 $\Omega_{\textrm{M0}}$\ \ & \ \ $0.269$\ \           & \ \ $0.269$\ \                 & \ \ $0.269$\ \                 & \ \ $0.269$\ \                 & \ \       $0.263$\ \           & \ \ $0.307$\ \  \\
 $\chi_{\rm min}^2$    \ \ & \ \ $494.530$\ \         & \ \ $448.726$\ \               & \ \ $504.496$\ \               & \ \ $458.857$\ \               & \ \ $455.554$\ \               & \ \ $452.839$\ \  \\
 $\chi_{\rm min}^2/\textrm{dof}$\ \ & \ \ $1.132$\ \           & \ \ $1.136$\ \                 & \ \ $1.160$\ \                 & \ \ $1.147$\ \                 & \ \ $1.145$\ \                 & \ \ $1.138$\ \  \\
 $\textrm{AIC}$        \ \ & \ \ $502.530$\ \         & \ \ $456.726$\ \               & \ \ $512.496$\ \               & \ \ $464.857$\ \               & \ \ $463.554$\ \               & \ \ $460.839$\ \  \\
$\textrm{BIC}$         \ \ & \ \ $518.877$\ \         & \ \ $472.672$\ \               & \ \ $528.825$\ \               & \ \ $482.853$\ \               & \ \ $479.530$\ \               & \ \ $476.815$\ \ \\
 \hline\hline
 \end{tabular} }
 \end{center}
 \caption{\label{tab2} The best-fit value of  the corresponding  ($w_{0}$, $w_{a}$)  with $1\sigma$ uncertainties
 and the best-fit value of $\Omega_{\textrm{M0}}$, and $\chi_{\rm min}^2$, $\chi_{\rm
min}^2/\textrm{dof}$, as well as $\textrm{AIC}$, $\textrm{BIC}$ for
the CPL model in a flat universe, with
357SNe+GRBs+others(CMB+BAO+CBF+$H$), 357SNe+others (without GRBs),
397SNe+others, and with 357SNe+GRBs+C(CMB)+B(BAO), 357SNe+GRBs+CMB,
and 357SNe+GRBs+BAO.}
 \end{table*}


 \begin{center}
 \begin{figure}[tbhp]
 \centering
 \includegraphics[width=0.45\textwidth]{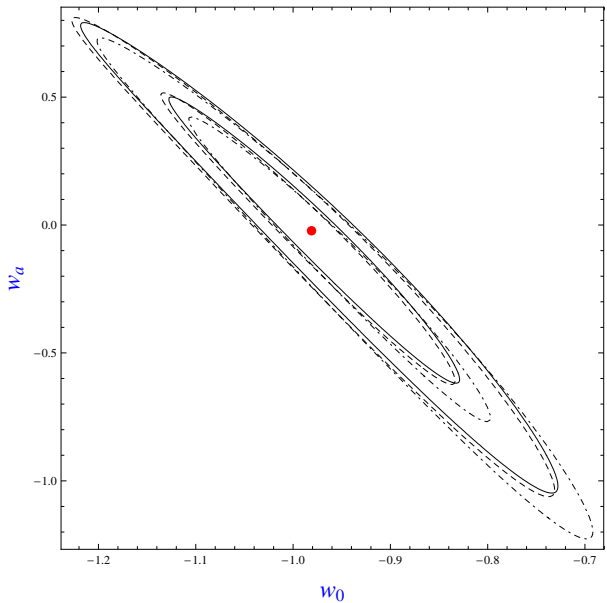}\hfill
 \caption{\label{figCPL1}
 The joint confidence regions in the
 ($w_0,w_a$) plane for the CPL model in a flat universe with 357SNe+GRBs+others(CMB+BAO+CBF+$H$),
 357SNe+others (without GRBs), and 397SNe+others. The contours correspond to
1-$\sigma$ and  2-$\sigma$ confidence regions, and the red point is
the best-fit value (357SNe+GRBs+CMB+BAO+CBF+$H$). The solid lines
represent the results of 357SNe+GRBs+others(CMB+BAO+CBF+$H$).  The
dash-dotted lines represent the results of  357SNe+others (without
GRBs). The dashed lines represent the results of 397SNe+others.}
 \end{figure}
 \end{center}


 \begin{center}
 \begin{figure}[tbhp]
 \centering
 \includegraphics[width=0.45\textwidth]{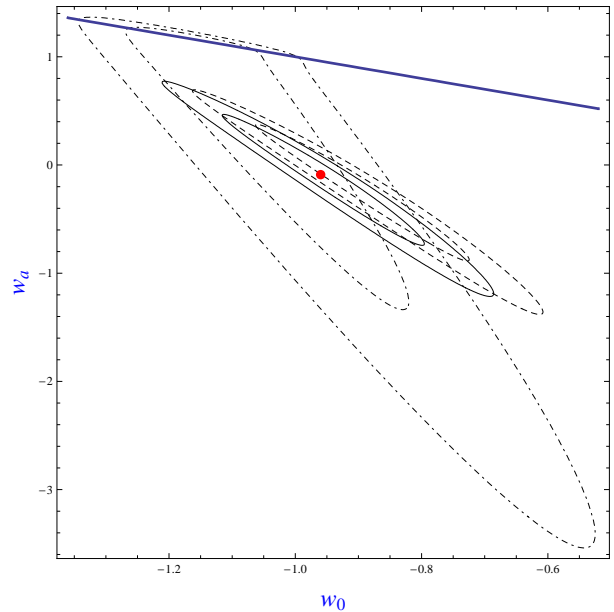}\hfill
 \caption{\label{figCPL2}
 The joint confidence regions in the
 ($w_0,w_a$) plane for the CPL model in a flat universe with
 357SNe+GRBs+CMB+BAO, 357SNe+GRBs+CMB and 357SNe+GRBs+BAO. The contours correspond
to 1-$\sigma$ and  2-$\sigma$ confidence regions, and the red point
is the best-fit value (357SNe+GRBs+CMB+BAO). The solid lines
represent the results of 357SNe+GRBs+CMB+BAO. The dashed lines
represent the results of 357SNe+GRBs+CMB. The dash-dotted  lines
represent the results of 357SNe+GRBs+BAO. The straight line near the
top is given by $w_a+w_0=0$.}
 \end{figure}
 \end{center}

For the flat CPL model, we find that the best-fit parameters with
357SNe + GRBs + CMB + BAO+ CBF + $H(z)$ are $(w_0, w_a,
\Omega_{\textrm{M0}})=(-0.98, -0.02, 0.269)$, with
$\chi^2_{\textrm{min}}= 494.53$ for 437 degrees of freedom. Figure 4
shows the joint confidence regions in the ($w_0,w_a$) plane, while
fixing $\Omega_{\textrm{M0}}=0.269$. For comparison, fitting results
from the joint data without GRBs are also given in Fig. 4. With
357SNe + CMB + BAO + CBF + $H(z)$, the best-fit values are $(w_0,
w_a, \Omega_{\textrm{M0}})=(-0.95, -0.14, 0.269)$, with
$\chi^2_{\textrm{min}}= 448.726/395$. Fitting results with 397SNe +
CMB + BAO + CBF + $H(z)$ are also given in Fig. 4 and the best-fit
values are $(w_0, w_a, \Omega_{\textrm{M0}})=(-0.99, 0.00, 0.269)$,
with $\chi^2_{\textrm{min}}= 504.496/435$. In addition, fitting
results from the joint data with 357SNe + GRBs + CMB + BAO, 357SNe +
GRBs + BAO, and 357SNe + GRBs + CMB  are given in Fig 5. We note
that the contours for 357SNe + GRBs + BAO are sharply cutoff at the
top, near the line given by $w_{0}+w_{a}=0$, as shown in the figure.
This is due to $w(z\gg1)<0$ or $w_{0}+w_{a}<0$ in the CPL
parameterization (i.e, the early universe is matter dominated)
implicitly required by the BAO data; this was also noted by Kowalski
et al \cite{sn307}. We present the best-fit value of the derived
$w_{0}$, $w_{a}$ with $1\sigma$ uncertainties, and the best-fit
value of $\Omega_{\textrm{M0}}$, $\chi_{\rm min}^2$, $\chi_{\rm
min}^2/\textrm{dof}$, as well as $\textrm{AIC}$, $\textrm{BIC}$ with
differently combined data set in Table 2.

Comparing to the joint constraints with GRBs and without GRBs, we
can see that the contribution of GRBs to the joint cosmological
constraints is a slight shift in the ($w_0,w_a$) plane to enclose
the $\Lambda$CDM model ($w_0=-1, w_a=-0$). From the joint data
[357SNe + CMB + BAO + CBF + $H(z)$] with GRBs for the flat CPL
model, we obtain $\Delta\textrm{BIC}=6.141$ with respect to the
$\Lambda$CDM model, indicating a strong preference for the
$\Lambda$CDM model; while from the joint data without GRBs for the
flat CPL model, we obtain $\Delta\textrm{BIC}=5.905$ with respect to
the flat $w$CDM model. Meanwhile comparing to Fig. 4 and 5, we find
the effects of adding the 26 $f_{\textrm{gas}}$ data and the 11
$H(z)$ data are not very significant, suggesting that the model
parameters are strongly constrained by SNe+GRBs+CMB+BAO. Comparing
to the joint constraints with SNe+GRBs+CMB and SNe+GRBs+BAO, we find
the confidence regions with SNe+GRBs+CMB seem to close to the joint
constraints with SNe+GRBs+CMB+BAO. It indicates that the
contribution of CMB data to the joint cosmological constraints
(SNe+GRBs) is more significant compared to that of BAO data.

From Fig. 2 - 5 and Table 1 - 2,  we can find that the $\Lambda$CDM
model is consistent with the joint data in the 1-$\sigma$ confidence
region. Comparing to the joint constraints with GRBs and without
GRBs, we can find the effect of GRBs to the joint cosmological
constraints, although the contribution of GRBs to the cosmological
constraints would not be  sufficiently significant compared to that
of SNe Ia at present. This is mainly caused by the relatively large
statistical scattering in the GRB relations and the relatively small
data set of GRBs compared to that of SNe Ia currently.


\section{Reconstructing the Acceleration History of the Universe}
Following a well-known procedure in the analysis of large scale
structure, Shafieloo et al  used a Gaussian smoothing function
rather than the top hat smoothing function to smooth the noise of
the SNe Ia data directly \cite{Shafieloo06,Shafieloo07}. In order to
obtain important information on interesting cosmological parameters
expediently, when doing the Gaussian smoothing, $\ln d_L(z)$, rather
than the luminosity distance $d_L(z)$ or distance modulus $\mu(z)$,
is studied by an iterative method \cite{Shafieloo06,Shafieloo07}.
Here we follow the iterative procedure and adopt results from Ref.
\cite{Wu08},
\begin{eqnarray}
\ln f(z)^s_n=\ln f(z)^s_{n-1}+N(z)\sum_i\big[\ln f^{obs}(z_i)-\ln
f(z_i)^s_{n-1}\big]N_i(z),
\end{eqnarray}
where $f(z)\equiv{D_L(z)/h}$,  $f(z)^s_n$ represents the smoothed
luminosity distance at any redshift $z$ after the $n$ th iteration,
$f(z)^s_{0}$ denotes a guess background model, and $ f^{obs}(z_i)$
is the observed one from the SNe Ia data, as well as the
normalization parameter $N(z)^{-1}=\sum_i N_i(z)$,
$N_i(z)=\exp\big[-\big(\ln^2\big((1+z)/(1+z_i)\big)\big)/\big(2\triangle^2\big)\big]$.
It has been shown that the results are not sensitive to the chosen
value of $\Delta$ and the assumed initial guess model. Here we use a
$w$CDM model with $w=-0.9$ and $\Omega_{\textrm{M0}}=0.28$ as the
guessed background model and we choose $\triangle=0.6$ \cite{Wu08}.

The best iterative result is obtained by minimizing
\begin{eqnarray}
\chi^2_n=\sum_i
(\mu(z_i)_n-\mu^{obs}(z_i))^2/\sigma^2_{\mu_{obs,i}}.
\end{eqnarray}
Once the $\chi^2_n$ reaches its minimum value, we stop the iterative
process and get the best result ($\chi^2_{\textrm{min}}$), with the
1-$\sigma$ uncertainties corresponding to
$\chi^2=\chi^2_{\textrm{min}}+1$. Figure 6 shows the computed
$\chi^2_n$ for the reconstructed results at each iteration for the
SN Ia and GRB data. We find that when at $n=16$ a minimum value of
$\chi^2_n$ is obtained.


 \begin{center}
 \begin{figure}[tbhp]
 \centering
 \includegraphics[width=0.45\textwidth]{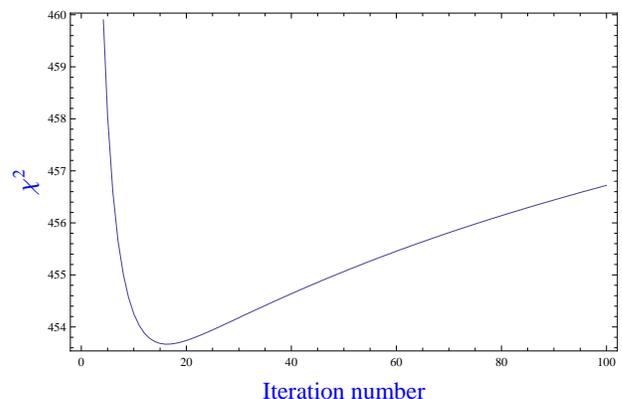}\hfill
 \caption{\label{fign}
 Computed $\chi^2_n$  for the reconstructed results at each
iteration for the SN Ia and GRB data. The minimum value of $\chi^2_n$ is obtained at $n=16$, and the 1-$\sigma$
uncertainties of reconstructed results can be obtained at $n=9, 41$, corresponding to $\Delta\chi^2=1$,
respectively.}
 \end{figure}
 \end{center}


By using the iterative approach to reconstruct the luminosity
distance at any redshift in the redshift range of SNe Ia, we have
calibrated GRB luminosity relations in a model independent manner
\cite{liang08b}. Here we reconstruct the Hubble parameter $H(z)$,
the deceleration parameter $q(z)$, and the \textsl{EoS} of dark
energy $w(z)$ from the best iterative result of $f(z)$ with the
distance moduli of SNe Ia and GRBs obtained by the interpolating
method.  The Hubble parameter can be given by differentiating the
smoothed luminosity distance \cite{Wu08},
\begin{eqnarray}
H(z)=\left\{\frac{d}{dz}\left[\frac{100f(z)}{(1+z)}\right]
 \right\}^{-1},
\end{eqnarray}
which contains the information on $H_0$. Then the deceleration
parameter $q(z)$  of  the expanding universe and the \textsl{EoS} of
dark energy can be obtained \cite{Wu08},
\begin{eqnarray}
q(z)=(1+z)\frac{H^\prime(z)}{H(z)}-1,
\end{eqnarray}
\begin{eqnarray}\label {calw}
w(z)=\frac{-1+\frac{2}{3}(1+z)\cdot
H'/H}{1-(1+z)^3\Omega_{\textrm{M0}}H_0^{2}/H^2},
\end{eqnarray}
where the prime ($\prime$) denotes the derivative with respect to
$z$.

 \begin{center}
 \begin{figure}[tbhp]
 \centering
 \includegraphics[width=0.45\textwidth]{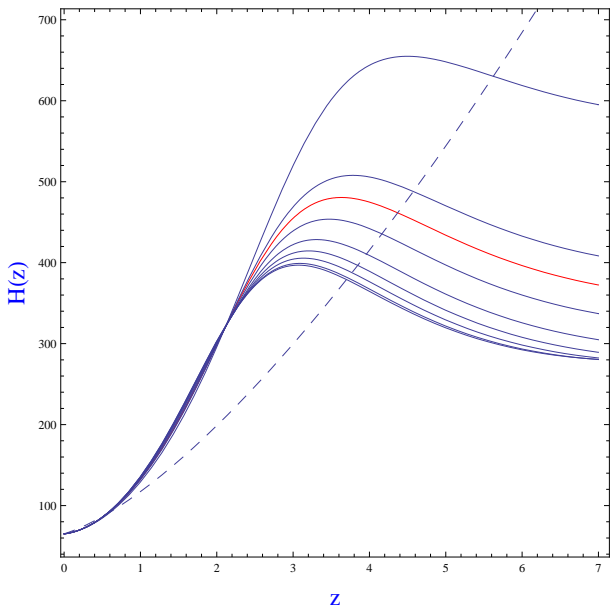}\hfill
 \caption{\label{figH}
 The reconstructed $H(z)$ with the 1-$\sigma$
uncertainties. The red line indicates the best-fit value of the
reconstructed $H(z)$ which is obtained at $n=16$ and the blue lines
represent the fit values of the reconstructed $H(z)$ within
1-$\sigma$ uncertainties at $n= 9, 14, 19, 24, 29, 34, 39, 41$. The
dashed line is the theoretical values of the $\Lambda$CDM model. }
 \end{figure}
 \end{center}



 \begin{center}
 \begin{figure}[tbhp]
 \centering
 \includegraphics[width=0.45\textwidth]{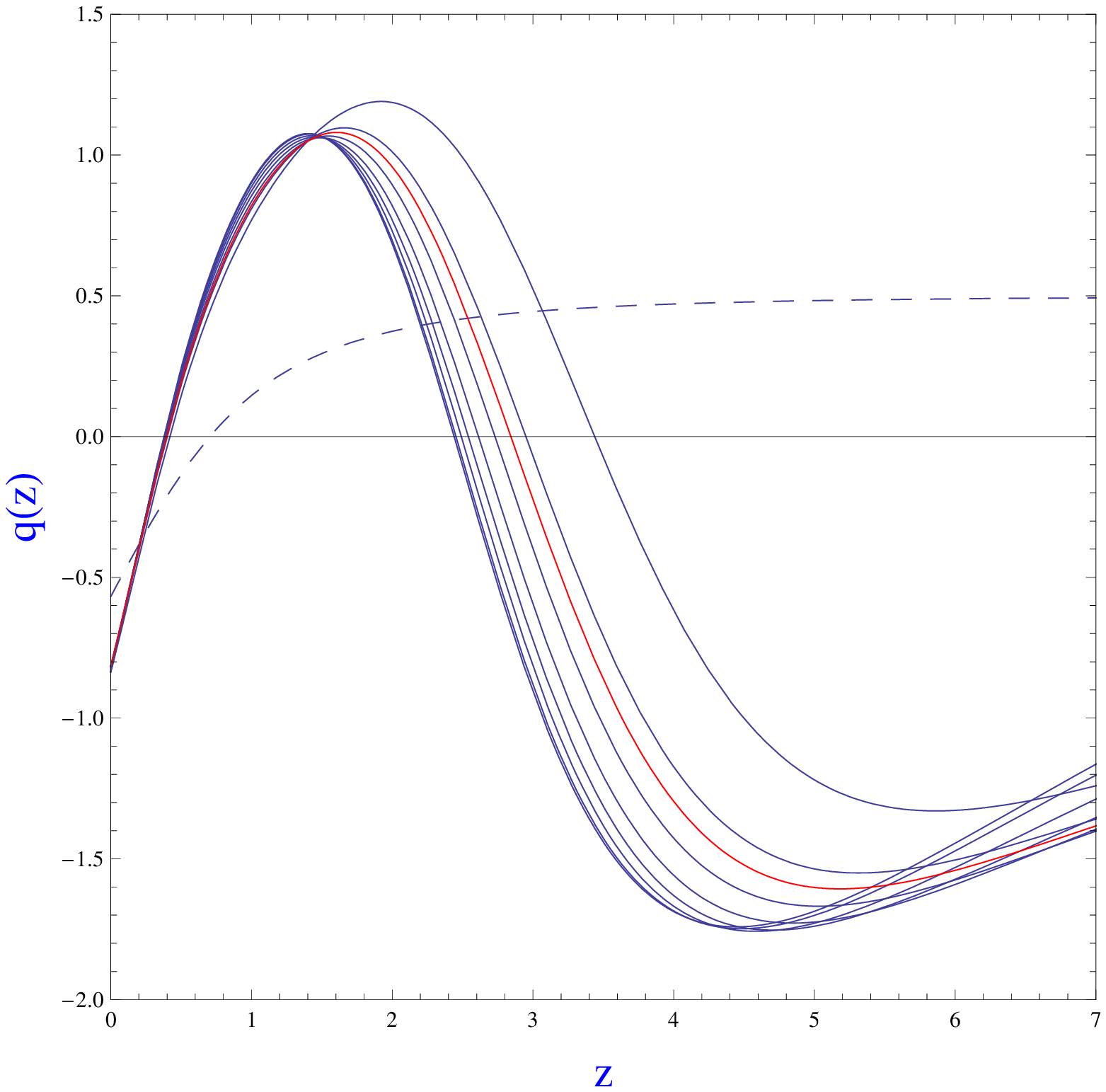}\hfill

 \caption{\label{figq}
The reconstructed $q(z)$ with the 1-$\sigma$ uncertainties. The red
line indicates the best-fit value of the reconstructed $q(z)$ which
is obtained at $n=16$ and the blue lines represent the fit values of
the reconstructed $q(z)$ within 1-$\sigma$ uncertainties at $n= 9,
14, 19, 24, 29, 34, 39, 41$. The dashed line is the theoretical
values of the $\Lambda$CDM model.}
 \end{figure}
 \end{center}



 \begin{center}
 \begin{figure}[tbhp]
 \centering
 \includegraphics[width=0.45\textwidth]{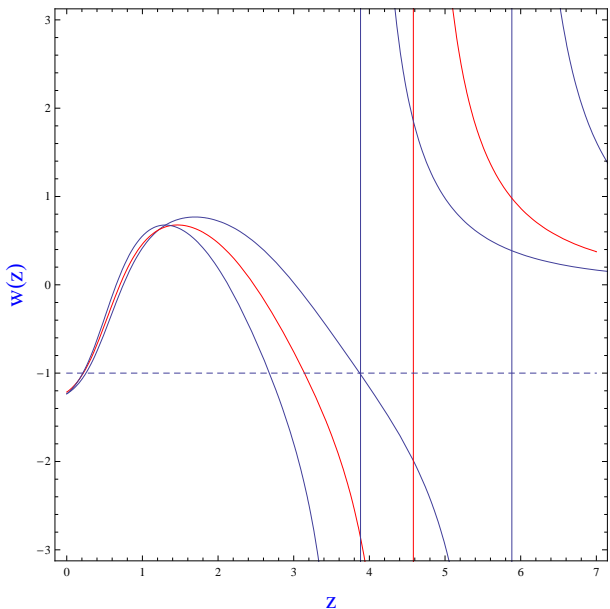}\hfill
 \caption{\label{figw}
The reconstructed \textsl{EoS} of dark energy $w(z)$ with the
1-$\sigma$ uncertainties. The red line indicates the best-fit value
of the reconstructed $w(z)$ which is obtained at $n=16$ and the blue
lines represent 1-$\sigma$ uncertainties at $n= 9, 41$. The dashed
line is the theoretical values of the $\Lambda$CDM model. }
 \end{figure}
 \end{center}

The reconstructed $H(z)$ with the 1-$\sigma$ uncertainties are shown
in Fig. 7. For comparison, the theoretical values of the
$\Lambda$CDM model are also shown in  Fig. 7. We can find that there
is a deviation between the reconstructed $H(z)$ and the theoretical
values of the $\Lambda$CDM model  with the 1-$\sigma$ uncertainty at
$1\lesssim z \lesssim4$, where the GRB data clearly dominate. The
reconstructed $q(z)$ with the 1-$\sigma$ uncertainties are shown in
Fig. 8. We can find that the transition redshift at which the
expansion of the universe from deceleration ($q(z)>0$) to
acceleration ($q(z)<0$) is $z_\textrm{T}=0.38^{+0.03}_{-0.03}$, at
relatively low redshifts, where the SNe Ia data dominate, which is
slightly later but more stringent than that reconstructed from the
ESSENCE supernova data ($z \sim 0.55-0.73$) \cite{Wu08}. However,
there is another transition redshift at which the expansion of the
universe from acceleration to deceleration,
$\tilde{z}_\textrm{T}\simeq3$, at high redshift where GRB data
dominate. However, this transition redshift has large 1-$\sigma$
uncertainties ($2.4<\tilde{z}_\textrm{T}<3.5$) compared to the
former one $0.35<z_\textrm{T}<0.41$, similar to that reconstructed
from the GRB data using the Amati relation \cite{wei09a}. The
reconstructed \textsl{EoS} of dark energy $w(z)$ with the 1-$\sigma$
uncertainties are shown in Fig. 9. We can find that there is a
singular point at $ z \lesssim5$ from the reconstructed $w(z)$. This
phenomenon has been known to happen and just displays the fact that
$w(z)$ [Eq 34]is nothing more than an effective parameter that (in
this case) fails to describe the system correctly. On the other
hand, the singular point may be caused by the absence of data in the
Hubble diagram at $5 \lesssim z \lesssim6$, i.e., between the GRB
data at $z<5$ and the only two GRBs at $z>6$ as shown in Fig. 1.

From Fig. 7 - 9, we can read that $H_0=66~\rm km\,s^{-1}\,Mpc^{-1}$,
$q_0=-0.82$ and $w_0=-1.19$. We can also find some features in the
reconstructed $H(z)$, $q(z)$, and $w(z)$, which seem to favor the
oscillating models \cite{oscillating1,oscillating2,oscillating3}.
However, it is noted that because the available data of GRBs at high
redshift are still quite rare now, and their statistical scatters
are relatively large compared to that of SNe Ia, these tentative
results might be artifacts, but nevertheless deserve further
investigations.


\section{Summary and Discussion}
Because of the lack of enough low red-shift GRBs to calibrate the
luminosity relation, GRBs could not be used reliably and extensively
in cosmology until recently. In our previous paper, we have proposed
a new method to calibrate GRB luminosity relations in a completely
cosmology-independent manner to avoid the well-known circularity
problem \cite{liang08a}.

In this work, with the recent GRB data at high redshift whose
distance moduli are calibrated with the interpolating method
\cite{liang08a} from the Union set of 307 SNe Ia \cite{sn307}, as
well as the Constitution set of SNe Ia \cite{sn397},the CMB
observation  from the WMAP5 result ($R$, $l_a$, $\Omega_bh^2$)
\cite{wmap5}, the BAO observation from the spectroscopic SDSS DR7
galaxy sample ($d_{0.2}$, $d_{0.35}$) \cite{SDSS7} , the x-ray
baryon mass fraction in clusters of galaxies \cite{Allen04}, and the
observational $H(z)$ data \cite{Simon05,Gazta08}, we find that the
$\Lambda$CDM model is consistent with the joint data in 1-$\sigma$
confidence region; this confirms the conclusion of many previous
investigations. We also find that the current GRB data are
substantially less accurate than the SNe Ia data, if each data set
is used alone, consistent with previous investigations.

The new results and insights we have obtained in this work are
briefly summarized as follows:

(1) In order to combine GRB data into the joint observational data
analysis to constrain cosmological models, we provide a simple
method to avoid any correlation between the SNe Ia data and the GRB
data; in this method those SNe Ia data points used for calibrating
the GRB data are not used.

(2) Comparing to the joint constraints with GRBs and without GRBs,
we find that the contribution of GRBs to the joint cosmological
constraints is a slight shift in the confidence regions of
cosmological parameters to better enclose the $\Lambda$CDM model.

(3) Finally we reconstruct the Hubble parameter $H(z)$, the
deceleration parameter $q(z)$, and the \textsl{EoS} of dark energy
$w(z)$ of the acceleration Universe up to $z>6$ with the distance
moduli of the Constitution set of SNe and GRBs and find some
features that seem to favor oscillatory cosmology models; however
further investigations are needed to better understand the
situation.

For considering the use of GRBs for cosmology, the gravitational
lensing effect may need to be considered \cite{Oguri06}. However
Schaefer found that the gravitational biases of GRBs are small
\cite{Sch07}. Recently, some possible observational selection bias
\cite{But07,Fiore07,Campana07,Sha09} and evolution effects
\cite{Li07,Tsu08,Salva08,Petro09} in GRB relations have also been
discussed. However,  Ghirlanda et al. confirmed the spectral-energy
relations of GRBs  observed by \emph{Swift} \cite{Ghi07}. Moreover,
it is found  that instrumental selection effects do not dominate for
the Amati relation \cite{Ghi08} and for the $E_{\rm peak}-L$
relation \cite{Ghi09}, as well as no strong evolution with redshift
of the Amati relation can be found  \cite{Ghi08,WDQ08}.
Nevertheless, for considering GRBs as standard candles to constrain
cosmology, further examinations of possible selection bias and
evolution effects should be required in a larger GRB sample.

Different dark energy models may have very different Hubble diagrams
at high redshifts \cite{Sch07}. Therefore, the best plan to
investigate the property of dark energy is measuring the dark energy
over a wide range of redshifts. GRBs can extend the Hubble diagram
to much higher redshifts beyond SNe Ia data \cite{Sch07}. It is
worth noticing that GRBs are important potential probes for cosmic
history up to $z>6$. However, the contribution of GRBs to the
cosmological constraints would not be sufficiently significant at
present, due to the large statistical scatters of the relations and
the small dataset of GRBs. As discussed in \cite{Sch07}, GRBs are
almost immune to dust extinction, whereas in the case of SNe Ia
observations, there is extinction from the interstellar medium when
optical photons propagate towards us. On the other hand, SNe Ia are
substantially more accurate standard candles than GRBs, which will
lead to tight constraints on cosmological parameters; whereas a
single GRB at high redshift will provide more information than a
single maximal redshift SNe Ia \cite{Sch07}. Through Monte Carlo
simulations, future prospect of probing dark energy parameters with
a larger sample of GRBs has been investigated
\cite{Taka03,Xu05,LZ05,Tsu09a}. It has been found that cosmological
constraints would improve substantially with more simulated GRBs
expected by future observations. Recently, Xiao and Schaefer have
compiled 107 long GRBs with their spectroscopic/photometric
redshifts measured \cite{XiaoSchaefer09}, observed by BATSE, Konus,
HETE, and \textit{Swift}. Along with more and more GRBs observed
from \emph{Fermi Gamma-ray Space Telescope} (formerly known as
\emph{GLAST}) with much smaller scatters, and its combination with
the increasing \emph{Swift} data, GRBs could be used as an
additional choice to set tighter constraints on cosmological
parameters of dark energy models.

\appendix
\section{The SN Ia data used  in the interpolating procedure for calibrating the GRB data }
 Here we list the 40 SNe Ia data points that  have been
used to obtain the distance moduli of the 27 GRBs at $z<1.4$ in the
interpolating procedure, and have consequently been excluded from
the SNe Ia sample used to the joint constraints.
\begin{table}[tbhp]{\scriptsize
\begin{tabular}{l|l|l|l|l|}
\hline \hline
&  SN           &   $z_{\textrm{SN}}$     &     GRB   &  $z_{\textrm{GRB}}$        \\
\hline
&    1999bp     &    0.1561     &   030329  &   0.17    \\
&    1996h      &    0.2130     &       &       \\
&    1995ao     &    0.2486     &   020903  &   0.25    \\
&    1995ap     &    0.2630     &       &       \\
&    2001hs     &    0.4300     &   990712  &   0.43    \\
&    2001fs     &    0.4300     &       &       \\
&    2001fo     &    0.4300     &       &       \\
&    2000fr     &    0.4300     &       &       \\
&    1998aw     &    0.4500     &   010921  &   0.45    \\
&    1998as     &    0.4500     &       &       \\
&    1997ez     &    0.4500     &       &       \\
&    04D3nc     &    0.6100     &   050525  &   0.61    \\
&    04D3ny     &    0.6450     &   050416  &   0.65    \\
&    04D3ml     &    0.6550     &       &       \\
&    03D4cn     &    0.6950     &   020405  &   0.70    \\
&    03D3aw     &    0.6950     &   970228  &   0.70    \\
&    03D1fl     &    0.7070     &   041006  &   0.71    \\
&    03D1cm     &    0.7100     &   991208  &   0.71    \\
&    d097   &    0.7800     &   030528  &   0.78    \\
&    e108   &    0.8000     &   051022  &   0.80    \\
&    f076   &    0.8300     &   050824  &   0.83    \\
&    f096   &    0.8300     &   970508  &   0.84    \\
&    f235   &    0.8400     &   990705  &   0.84    \\
&    f244   &    0.8400     &   000210  &   0.85    \\
&    f308   &    0.8540     &       &       \\
&    g050   &    0.8600     &   040924  &   0.86    \\
&    h364   &    0.9600     &   970828  &   0.96    \\
&    k411   &    0.9700     &   980703  &   0.97    \\
&    k425   &    0.9700     &       &       \\
&    k485   &    1.0100     &   021211  &   1.01    \\
&    m027   &    1.0100     &       &       \\
&    m062   &    1.0200     &   991216  &   1.02    \\
&    m138   &    1.0200     &       &       \\
&    m193   &    1.0570     &   000911  &   1.06    \\
&    m226   &    1.1200     &   980631  &   1.10    \\
&    n278   &    1.2300     &   050408  &   1.24    \\
&    n285   &    1.2650     &   020813  &   1.25    \\
&    n326   &    1.3000     &   050126  &   1.29    \\
&    p454   &    1.3050     &   990506  &   1.31    \\
&    p455   &    1.3400     &       &       \\
\hline
&  Number(SN) &    40     &     Number(GRB)   &  27        \\

\hline\hline
\end{tabular}
\caption{The SNe Ia data points used  in the interpolating procedure
for calibrating the GRB data at $z<1.4$. Columns 1 and 2 are SNe Ia
used  in the interpolating procedure with their redshifts; Columns 3
and 4 are the GRB data at $z<1.4$ with their redshifts.}}
\end{table}

\begin{acknowledgments}
We thank the anonymous referee for comments and suggestions. We also
thank Professors Zong-Hong Zhu, Ti-Pei Li, Rong-Gen~Cai, Zi-Gao Dai,
Ren-Cheng Shang, Zhan Xu, Fang-Jun Lu, En-Wei Liang, and Hongwei Yu
for kind help and discussions. N.L. thanks Yun Chen, He Gao, Shuo
Cao, Hao Wang, Fang Huang, Jie Ma, Xingjiang Zhu, and Dr. Yi Zhang
for discussions. P.X.W. acknowledges partial supports by the
National Natural Science Foundation of China under Grant No.
10705055, the Program for NCET (No. 09-0144), and the FANEDD under
Grant No.200922. S.N.Z. acknowledges partial funding supports by
Directional Research Project of the Chinese Academy of Sciences
under Project No. KJCX2-YW-T03 and by the National Natural Science
Foundation of China under Grant No. 10821061, No. 10733010, No.
10725313, and by 973 Program of China under Grant No. 2009CB824800.
\end{acknowledgments}


\end{document}